# Electrostatic effect on off-field ferroelectric hysteresis loop in piezoresponse force microscopy

*Huimin Qiao[1], Owoong Kwon[1,2], Yunseok Kim[1,2,*]*

[1]School of Advanced Materials and Engineering and [2]Research Center for Advanced Materials Technology, Sungkyunkwan University (SKKU), Suwon 16419, Republic of Korea

*Electronic mail: yunseokkim@skku.edu

# ABSTRACT

Piezoresponse force microscopy (PFM) has been extensively utilized as a versatile and an indispensable tool to understand and analyze nanoscale ferro-/piezo-electric properties by detecting the local electromechanical response on a sample surface. However, it has been discovered that the electromechanical response not only originates from piezoelectricity but also from other factors such as the electrostatic effect. In this study, we explore the dependence of off-field PFM hysteresis loops on the surface-potential-induced electrostatic effect in a prototypical ferroelectric thin film by applying an external voltage to the bottom electrode during measurement. We simplify the situation by equating the surface potential to the direct current voltage waveform variations and predicting the contribution of the surface-potential-induced electrostatic effect to the PFM hysteresis loops. The experimental results approximately match our prediction—the coercive voltage linearly decreases with the surface potential, whereas the saturated amplitude and piezoresponse remain nearly constant owing to the relatively large piezoelectric coefficient of the ferroelectric thin film.

Atomic force microscopy (AFM) is a versatile tool for characterizing high-resolution nanoscale surface properties.[1-4] A prominent example of this technique is piezoresponse force microscopy (PFM), which is mainly applied for the detection and manipulation of ferroelectric domains.[5-8] PFM is based on the detection of the dynamic electromechanical response originating from the converse piezoelectricity of piezoelectric and ferroelectric materials when an alternating current (AC) voltage is applied to the AFM tip.[9-11] Furthermore, the local ferroelectric switching properties can be obtained from hysteresis loop measurements. The high resolution down to the nanoscale and non-destructive features make the well-established PFM technique an indispensable tool for characterizing ferroelectric ultrathin films and nano-sized devices.[12-16]

The local hysteresis loop has been considered an evidence of the ferroelectric nature of materials, and it can also be used to evaluate the piezoelectric coefficient, especially when macroscopic electric measurements are difficult to perform.[17,18] However, beyond piezoelectricity, the electromechanical strain on the surface can also stem from alternative mechanisms such as Joule heating,[19] ion motion,[20] electrochemical strain,[21] and the electrostatic effect.[22] The electrostatic effect is the most ubiquitous factor among these because it is not limited to specific materials;[23] it is related to the electrostatic potential difference between the AFM tip/cantilever and the sample surface. Additionally, charge injection by the AFM tip voltage can induce the electrostatic effect and influence the electromechanical response.[24] For example, analogous to ferroelectricity, the electrostatic interaction can produce electromechanical hysteresis.[24-26] To accurately interpret the electromechanical response in a hysteresis loop, it is necessary to gain an insight into the electrostatic contributions to the PFM signal. It has been reported that the electrostatic effect can be eliminated by using a stiff tip or applying a nullifying voltage.[27] In contrast, a voltage waveform consisting of a pulsed triangular DC voltage and a continuous sinusoidal AC voltage has been employed to measure both on-field and off-field loops.[23, 28] The off-field loop, which records the remnant piezoresponse signal, is assumed to solely contain the piezoelectricity-induced electromechanical response, whereas the on-field loop is a mixture of piezoelectricity and the electrostatic effect from the DC voltage sweeping. Therefore, the off-field loop is more suitable for analyzing the electromechanical response from piezoelectricity. However, nearly all previous studies have focused on the electrostatic effect in an extreme case of the hysteresis loop such as on-field hysteresis loops or on the images in classical ferroelectrics.[22, 29] Although the electrostatic effect on off-field hysteresis loops was recently reported in piezoelectrics (without ferroelectricity) and dielectrics (without piezoelectricity),[23, 30] information concerning the influence of electrostatic effects on the off-field hysteresis loop in ferroelectrics is scarce.

In this study, we investigate the electrostatic contribution to the off-field ferroelectric hysteresis loops in a $BiFeO_3$ (BFO) thin film by varying the measurement conditions, such as the maximum DC voltage ($V_{DC,max}$) of the triangular waveform and the external DC voltage ($V_{DC,BE}$) of the bottom electrode, during the hysteresis loop measurement. The electrostatic effect induced by the DC voltage sweeping during the measurement on the PFM signal is found to be marginal. However, the electrostatic effect related to the external DC voltage significantly affects the hysteresis loop.

A rhombohedral BFO thin film with an approximate thickness of 50 nm were used in this study. The hysteresis loop and surface-potential measurements were recorded using a commercial AFM (NX10, Park Systems). A Pt/Ir-coated conductive AFM tip with a nominal spring constant of 3 N/m (Multi75E-G) was used. Details on sample preparation and the experiment setup and conditions can be found in the supporting information. The PFM amplitude was calibrated using inverse optical laser sensitivity that was extracted from the force-distance curve.[23] We note that the measurements were performed at a single point to avoid the influence of inhomogeneity; furthermore, the measurements for each condition in Figs. 1 and 3 were performed for twice to minimize additional electrostatic effects related to charge injection; the second loop is presented in the figures.

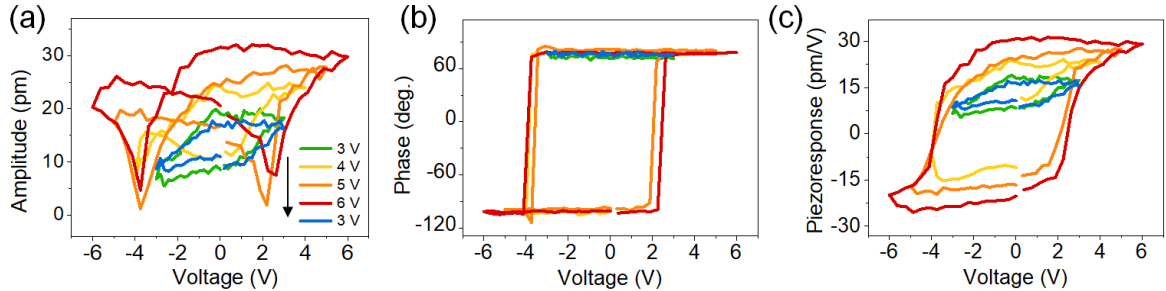

Fig. 1. Off-field (a) amplitude, (b) phase, and (c) piezoresponse hysteresis loops of the BFO sample with various $V_{DC,max}$ values at a single point. The black arrow indicates the measurement sequence.

Prior to the evaluation of the electrostatic effect on off-field hysteresis loops, we first analyze the hysteresis loop measurements with various values of $V_{DC,max}$, as presented in Fig. 1, to understand the switching behavior. At a low voltage of 3 V, the single wings in the amplitude hysteresis loop (Fig. 1a) and unchanged phase (Fig. 1b) indicate that the applied $V_{DC,max}$ is not sufficient to induce a full polarization reversal along the thickness direction. After increasing the voltage up to 4 V, two asymmetrical wings are observed in the amplitude hysteresis loop; the wing on the left side exhibits a sharp valley, which indicates a polarization reversal, whereas the one on the right side is flat. Additionally, the signs of phase and piezoresponse change on the left side, indicating that a polarization reversal occurs when a negative voltage of –4 V is applied to the tip. The asymmetrical polarization switching may be linked with the presence of a built-in electric field caused by the different top and bottom electrodes.[31, 32] Further increases in the tip voltage (5 V and 6 V) lead to a butterfly-shaped amplitude hysteresis loop with two sharp switching valleys; full polarization switching is observed on both left and right branches of the amplitude, phase, and piezoresponse hysteresis loops. Then, we repeatedly measure the hysteresis loop with a $V_{DC,max}$ of 3 V; the amplitude signal exhibits a slight change, compared to the one we obtained from the pristine state. This may be relevant to the non-significant electrostatic effects that will be discussed later in connection with Fig. 3.[22, 23]

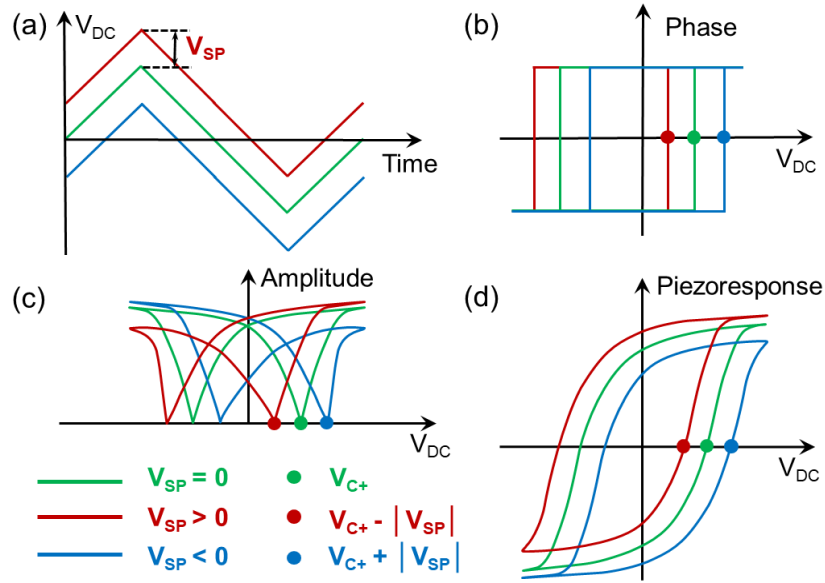

Fig. 2. Schematics of (a) effective excitation voltages applied to the tip ($V_{DC}$), and (b) phase, (c) amplitude, and (d) piezoresponse loops with different surface potentials ($V_{SP}$).

In general, accurate determination of the surface potential induced by the application of an external DC voltage is difficult because of the fast dissipation of injected charges after the DC voltage is turned off.[30] However, the electrostatic contribution from an external DC voltage to the bottom electrode can have an analogous influence on the PFM signal as a surface potential with the opposite sign.[12, 19] In this study, we investigate the electrostatic effect of the surface potential on the hysteresis loops using an analogous approach. By applying an external DC voltage ($V_{DC,BE}$) to the bottom electrode of the sample, we can correspondingly modulate the surface potential. Similar to our previous study,[30] the surface potential ($V_{SP}$) is assumed to be the same as $-V_{DC,BE}$ in the present case. However, if the BFO thin films can exhibit high conductivity, this assumption must be reconsidered. Then, we simplify the analysis by equating the surface-potential variation to the change in the waveform of the DC sweeping voltage, as schematically presented in Fig. 2a. The effective DC sweeping voltage integrally moves upward (downward) to nullify the positive (negative) surface potential.[36] Consequently, the critical voltage parameters (i.e., the coercive voltages $V_{C-}$ and $V_{C+}$) change accordingly, and the variations in the hysteresis loops along the *x* axis (voltage) can be predicted (Figs. 2b–2d). When a positive surface potential exists, the coercive voltage negatively shifts by the value of the positive potential, which can be written as $V_{C+} - |V_{SP}|$; similarly, a negative surface potential leads to a positive shift of $V_{C+} + |V_{SP}|$. This indicates that incorrect coercive voltages can be obtained because of the electrostatic effects. In addition to the coercive voltage, the amplitude can also be changed by the electrostatic effect.[27, 30] The first harmonic amplitude of the PFM signal can be described as[12]

$$A_{\omega,z} = d_{zz}V_0\sin(\omega t) + \frac{1}{k}\frac{dC}{dz}(V_{DC} - V_{SP})V_0\sin(\omega t), \tag{1}$$

where $d_{zz}$, $k$, $C$, and $V_0$ are the piezoelectric coefficient, contact stiffness of the cantilever, and capacitances of the tip and surface junction, respectively. For the off-field case, because $V_{DC}$ is zero, Eq. (1) can be modified as

$$A_{\omega,z} = d_{zz}V_0\sin(\omega t) - \frac{1}{k}\frac{dC}{dz}V_{SP}V_0\sin(\omega t). \tag{2}$$

Here, the capacitance derivative is negative,[12] which implies that the amplitude of the PFM signal varies with $V_{SP}$ by a ratio of $-\frac{1}{k}\frac{dC}{dz}V_0\sin(\omega t)$, which can also be referred to as the electrostatic coefficient. Consequently, the amplitude and corresponding piezoresponse hysteresis loops not only shift along the $x$ axis but also along the $y$ axis, as depicted in Figs. 2c and 2d.

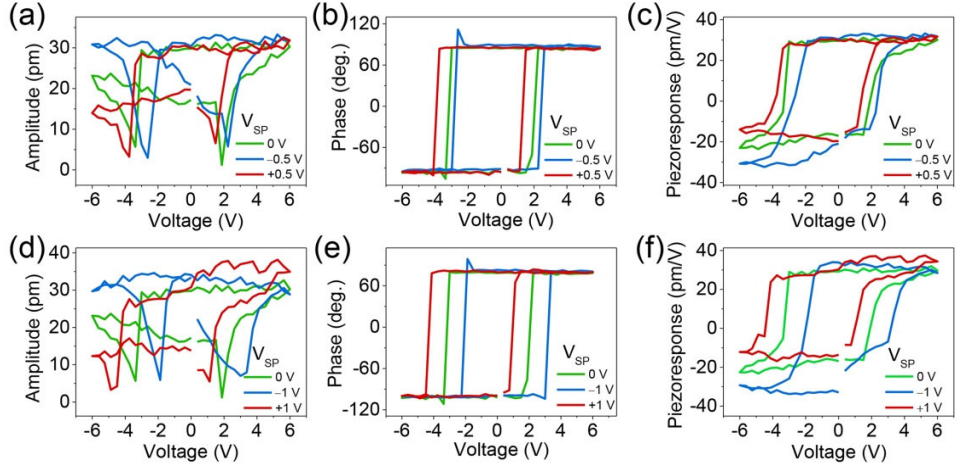

Fig. 3. Off-field (a),(d) amplitude; (b),(e) phase; (c),(f) piezoresponse hysteresis loops of the BFO sample with different surface potentials ($V_{SP} = -V_{DC,BE}$); (a-c) −0.5 V, 0 V, +0.5 V and (d-f) −1 V, 0 V, +1 V.

To examine our prediction described above, we obtained the hysteresis loops with different surface potentials by varying $V_{DC,BE}$. In this experiment, we used a $V_{DC,max}$ of 6 V because this is the critical voltage to obtain saturated loops based on the results in Fig. 1. As expected, the loops and coercive voltages shift negatively (positively) in the presence of a positive (negative) surface potential (Fig. 3). The positive and negative branches of the amplitude and piezoresponse hysteresis loops exhibit different behaviors in the presence of a surface potential; thus, we analyze them separately. For the positive branch, we observe that the difference in amplitude at $V_{DC,max}$ is marginal in the presence of a positive or negative surface potential. Since the saturated state is reached for the positive branch, only the electrostatic effect from the surface potential, as described in Eq. 2, contributes to the variation in the amplitude and piezoresponse signals. Therefore, the marginal change in amplitude may be because the electrostatic coefficient is so small that the second term in Eq. 2 is negligible in comparison with the first term even with a surface potential of ±1 V. In Ref. 30, the surface-potential-induced electrostatic contribution to the amplitude was significant because the electrostatic coefficient (0.213 pm/V) is comparable with the $d_{zz}$ (1 pm/V) of quartz. For a BFO, if we assume that the electrostatic coefficient is similar to that in the case of quartz because the model of cantilevers used is the same, then the $d_{zz}$ of BFO (~30 pm/V) would be much larger than the electrostatic coefficient; hence, the contribution of the surface potential to the amplitude and piezoresponse signals in a BFO is not as significant as it is in the case of quartz. For the negative branch, the considerable variations in amplitude and piezoresponse signals are assumed to be induced by the shift in the *x* axis, i.e., the shift in the negative coercive voltage. All the measurements were performed at the same position to preclude the influence of inhomogeneity of the sample. The surface potentials after each measurement were recorded to explore additional electrostatic effects related to the charge injection, as illustrated in Fig. S2.[33-35] The average surface potential decreases (increases) slightly after a positive (negative) $V_{DC,BE}$ is applied. However, the value of the surface potential is considerably smaller than that of the applied voltage, which may be a result of fast diffusion or compensation for the injected charges; furthermore, the change in the surface potential is still within the error range. Moreover, although the values of amplitude and coercive voltage for the additional measurements at different locations in Fig. S3 are slightly different from the ones in Fig. 3, the same tendency of variations in amplitude and coercive voltage with surface potential proves the validity and repeatability of our methods. We note that, although ionic effects can exist in a BFO,[37] no visible static deformation was observed after the hysteresis loops (not shown here) in this study. Thus, the electrostatic effect can be a major contribution to the change in the PFM signal.

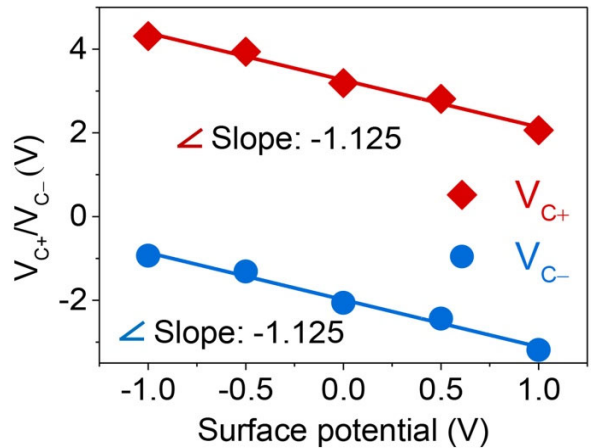

Fig. 4. Variations in positive and negative coercive voltages ($V_{C+}$ and $V_{C-}$) with surface potential; the data were extracted from Fig. 3. The solid lines are the fitting results.

Because we assume that the coercive voltage varies with $V_{SP}$ when a surface potential exists, the coercive voltage is expected to decrease with the surface potential by a slope of −1. Therefore, we extracted the coercive voltages (both positive $V_{C+}$ and negative $V_{C-}$) from each loop in Fig. 3 and plotted them as a function of the surface potential, $V_{DC,BE}$. The value of $V_{C+}$ ($V_{C-}$) is

+3.19 V (–2.06 V) when $V_{DC,BE}$ is not applied and where the surface potential is approximately zero, as presented in Fig. S2a. The increase in surface potential to +0.5 V and +1 V induces a decrease in $V_{C+}$ ($V_{C-}$) to 2.81 V (–2.44 V) and 2.06 V (–3.19 V), respectively. The decrease in surface potential to –0.5 V and –1 V induces an increase in $V_{C+}$ ($V_{C-}$) to 3.94 V (–1.31 V) and 4.31 V (–0.94 V), respectively. In Fig. 4, we linearly fitted the variation in coercive fields as a function of surface potential and obtained a parameter of –1.125, which is reasonably close to the expected value of –1. We also extracted and fitted the coercive fields extracted from the loops presented in Figs. S3a–S3c; all the parameters resulting from the fittings are close to –1 as presented in Figs. S3e and S3f. The marginal deviation may be attributed to the slight surface-potential variations prior to the hysteresis loop measurement. Overall, the electrostatic effect can shift the hysteresis loops and coercive voltages in a certain direction depending on the sign of the surface potential. This indicates that if the electrostatic effect is a significant part of the PFM signal, it is difficult to obtain the intrinsic coercive voltages of ferroelectric materials. We note that, because the same method was used to effectively investigate the electrostatic effect in the case of quartz,[30] which is a piezoelectric material with a low $d_{zz}$, we can conclude that our method provides a universal approach to evaluating the electrostatic effect in piezoelectric and ferroelectric materials.

In summary, we studied the influence of the electrostatic effect on off-field hysteresis loops in BFO thin films by applying external DC voltages to modulate the surface potential. The experimental results indicate that the variations in the saturated amplitude and piezoresponse with respect to the surface potential were negligible, which may be attributed to the relatively small electrostatic coefficient and large piezoelectric coefficient of the BFO. Meanwhile, a linear decrease in the coercive voltage with an increase in the surface potential, with a slope of approximately –1, was observed, which is close to our assumption. Thus, our work is expected to provide a basis for understanding the electrostatic effect of surface potential on PFM hysteresis loops in ferroelectric materials.

## Supplementary Material

See the supplementary material for the sample and experimental details, surface potential images, and repeated results.

This work was supported by a National Research Foundation of Korea (NRF) grant funded by the Korean government (MSIP; No. 2019R1I1A1A01063888) and the Basic Science Research Program through the National Research Foundation of Korea (NRF) funded by the Ministry of Education (No. 2019R1A6A1A03033215).

## Data Availability Statement

The data that support the findings of this study are available from the corresponding author upon reasonable request.

# Supplementary Material

# Electrostatic effect on off-field ferroelectric hysteresis loop in piezoresponse force microscopy

*Huimin Qiao[1], Owoong Kwon[1,2], Yunseok Kim[1,2,*]*

[1]School of Advanced Materials and Engineering and [2]Research Center for Advanced Materials Technology, Sungkyunkwan University (SKKU), Suwon 16419, Republic of Korea

*Electronic mail: yunseokkim@skku.edu

# 1. Experimental methods

The BFO thin film was epitaxially grown on an ~10 nm-thick $SrRuO_3$-buffered (001) $SrTiO_3$ substrate via pulsed laser deposition. After deposition, the sample was slowly cooled to ~25 °C in an oxygen environment of 500 Torr to minimize the formation of oxygen vacancies. The PFM hysteresis loops were collected using an AFM equipped with a function generator and data acquisition system (PXI-5412/5122, National Instruments), which were controlled with a custom software developed in LabVIEW/MATLAB. A band excitation waveform of 1 $V_{pp}$ in the range of 320–400 kHz was applied during measurement. A pulsed triangular voltage waveform (see Fig. S1) was used to acquire the off-field hysteresis loop. The surface potential was measured using amplitude-modulated Kelvin probe force microscopy. The experiments were conducted in ambient conditions with a temperature of ~28 ºC and humidity below 10%.

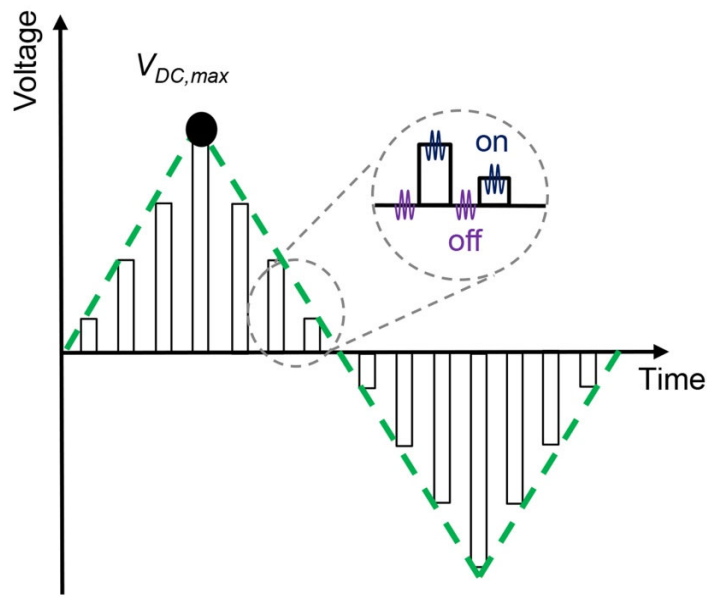

Fig. S1. Schematic of the voltage waveform.

## 2. Surface potential

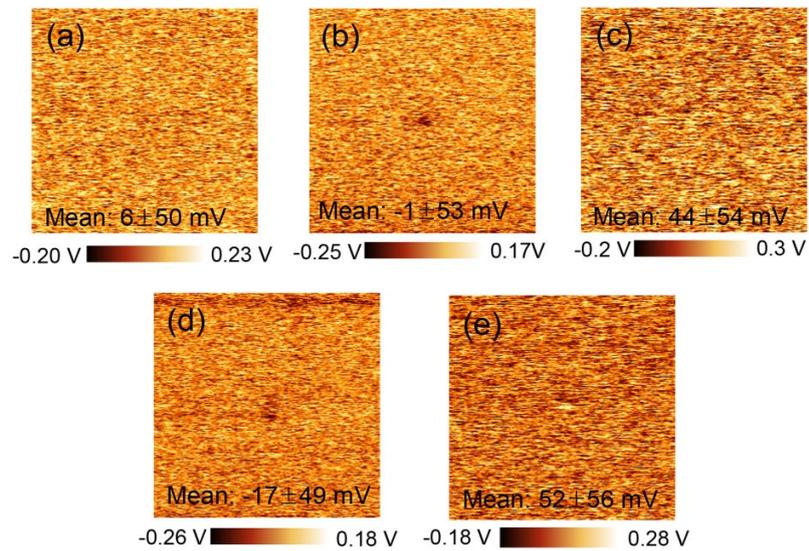

Fig. S2. Surface potential images (a) the as-grown state, and (b-e) after hysteresis loop measurement with sample biases of (b) +0.5 V, (c) −0.5 V, (d) +1 V, and (e) −1 V. The image size is 2 × 2 μm.

Although slight changes in the surface potential were observed after the application of the positive sample biases, the surface potential variation is so small in comparison with the deviation that it can be regarded as within the error range.

## 3. Repeated experimental measurements of hysteresis loop

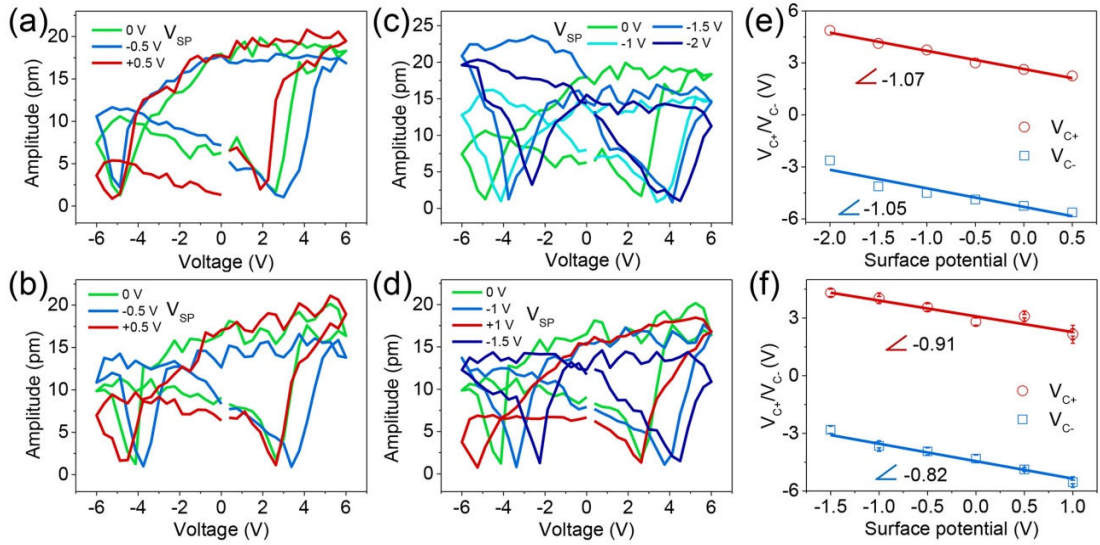

Fig. S3. Repetition of the results shown in Fig. 3 for two different locations: (a, c, e) location 1, (b, d, f) location 2. (a-d) Amplitude hysteresis loops at different surface potentials, (e, f) variation of positive and negative coercive fields with surface potential; the solid lines denote fitting results.

Figure S3 presents the results of the repetition of the same experiment illustrated in Fig. 3 at two different locations. For location 1, two cycles of the waveform illustrated in Fig. S1 were used during the measurement, and the values in Fig. S3e were extracted from the second loops in Figs. S3a and S3c. For location 2, five cycles of the waveform were used, and the loops presented in Fig. S3b and S3d are the fifth loops at each surface potential; the values in Fig. S3f are averaged from the second to fifth loops, and the error bar is used to show the variation in each loop. The saturation degree of the loops in Fig. S3 is not as high as that of the loops in Fig. 3, which may be one of the reasons for the deviation in the amplitude. On the other hand, inhomogeneities such as domain switching, built-in polarization, and/or surface states can contribute to the differences in the amplitude and coercive field. The slopes in Fig. S3f are slightly lower than those in Fig. S3e. This could be correlated with the fact that the injected charges during voltage sweeping cannot be released in a timely fashion and hence influence the next hysteresis loop because there is insufficient time for charge release after each waveform is completed; however, it could also be related to the homogeneities.